\documentclass[twocolumn,aps,showpacs]{revtex4-1}

\usepackage{amsmath}
\usepackage{graphicx}
\usepackage{bm}
\usepackage[breaklinks=true]{hyperref}
\usepackage[titletoc]{appendix}

\input epsf.sty

\begin{document}

\title{Influence of random pinning on the crystallization process in suspensions of hard spheres} 

\author{Sven Dorosz}
\affiliation{Theory of Soft Condensed Matter, Universit\'e du Luxembourg, L-1511 Luxembourg, Luxembourg}

\author{Tanja Schilling}
\affiliation{Theory of Soft Condensed Matter, Universit\'e du Luxembourg, L-1511 Luxembourg, Luxembourg}

\begin{abstract}
We discuss crystal formation in supersaturated suspensions of monodisperse hard spheres with a concentration of hard spheres randomly pinned in space and time. 
The pinning procedure introduces an external length scale and an external time scale that restrict the accessible number of configurations and ultimately the number of pathways leading to crystallization. We observe a significant drop in the nucleation rate density at a characteristic pinning concentration that can be directly related to the structure of the critical nucleus and the dynamics of its formation in the unpinned system. 
\end{abstract}

\maketitle

\section*{Introduction}
Homogeneous as well as heterogeneous crystallization are of importance in materials design and production. But even for one of the most simple models for liquids, the suspension of monodisperse hard spheres, the crystallization process is not fully understood \cite{Lechner2011, Schilling2010, Tanaka2013, Pusey2009, Voivod2009, Russo2013,Auer2001,Filion2010,TaSv2011}. \\
For the hard sphere system the transition from the supersaturated fluid to the crystal is purely entropic. It is a first order transition, hence in the case of packing fractions slightly higher than the coexistence packing fraction the system prevails in its meta-stable fluid state for a characteristic induction time before it is transformed irreversibly into a crystal. \\
The idea of the present work is to modify and restrict the possible number of pathways to crystallization in a controlled manner to understand how sensitive the crystallization process and, in particular, the induction time are with respect to changes in configuration space. \\
The method we employ is to take a configuration of hard spheres and to pin a randomly chosen fraction of them to their current positions. This model is called the random pinning model (RP) in the literature. Its diffusive behavior and relaxation properties have already been studied numerically and analytically (see \cite{Kim2003,Viramontes-Gambo1995,Krakoviak2010, Kurzidim2010, Kurzidim2011, Kim2011, Kim2003} and references therein). For high pinning concentrations, the relaxation times become large and the dynamic scattering function displays two step relaxation characteristic of glassy dynamics \cite{Kim2003,Kurzidim2010}.\\
Even though the dynamics change, the configurations correspond to typical equilibrated fluid configurations \cite{Krakoviak2010}. Therefore, static properties, as for example the static structure factor or the pair correlation function do not display any signatures of the restricted number of configurations.\\ In case the system undergoes the phase transition to the crystal, though, the transition process is modified.\\
Recently, computer simulations of the RP model were used to detect signatures of a static length scale associated with the glass transition in systems of hard spheres \cite{Scheidler2002, Berthier2012, Berthier2011, Kim2003, Krakoviak2005, Krakoviak2011} and mode coupling theory \cite{Cammarota2012, Lang2010}.\\ 
In the present work, the pinning concentrations are kept sufficiently low not to reach the glass transition, because we are interested in crystallizing trajectories.
Our study is split in two parts: In the first part we discuss crystallization for static random pinning, i.e.~once the pinned hard spheres are chosen, they stay pinned for the rest of the simulation. The concentration of pinned hard spheres can directly be translated into a length scale that interferes with the typical size of a critical cluster. In the second part we alter the selection of pinned hard spheres in time intervals $\Delta T$. $\Delta T$ directly interferes with the typical time scale to form a critical cluster. Due to the external time scale $\Delta T$, trajectories in configuration space are only restricted temporarily. The frustration due to the pinned hard spheres on long time scales is resolved. \\

\subsection*{Simulation Method}
We investigate $N=216000$ hard spheres of diameter $\sigma=1$ at constant volume V and constant energy E. We focus on the packing fraction $\eta = 0.5393$ which corresponds to a chemical potential difference between the metastable liquid and the stable crystalline state of $\Delta \mu \simeq -0.54\;k_BT$ at zero pinning concentration. The chemical potential difference has been obtained by integrating along the metastable fluid branch and the stable crystal branch of the equation of state.\\
The time evolution of the system is calculated using an event driven molecular dynamics algorithm (EDMD), see \cite{Alder1959,Lubachevsky1991}. Periodic boundary conditions are applied in all three directions of space. The initial velocities are drawn from a Gaussian distribution and the mean kinetic energy per hard sphere is set to $3k_BT$. A fraction $c$ of hard spheres is chosen randomly and pinned in space. They effectively possess zero velocity and infinite mass.\\ 
We first discuss static pinning. Here, a set of hard spheres of concentration $c\in \{0.00001, 0.0001, \ldots 0.1\}$, is chosen at the beginning of the simulation and pinned throughout.
Then we continue with periodic pinning. After time intervals $\Delta T$, a new set of hard spheres is randomly pinned and for the other hard spheres new velocities are randomly chosen from a Gaussian distribution. Here, we will focus on $c=0.05$ because in the case of static pinning, $c=0.05$ is the largest concentration that allows us to observe crystallization.  $\tau= 1\sqrt{\frac{m\sigma^2}{k_BT}}$ is the natural time unit of the simulation algorithm. The pinning time intervals discussed are $\Delta T\in \{0.01\tau, 0.02\tau, 0.05\tau, \ldots 16\tau, 32\tau,$ and $ 64\tau \}$. \\
During the molecular dynamics simulation the local $q_6q_6$-bond order parameter \cite{Steinhardt1983,tenWolde1995} is evaluated to monitor the size of the largest crystalline cluster.\\  
For a hard sphere $i$ with $n(i)$ neighbors (satisfying $r_{ij} < 1.4\sigma$) the local orientation is characterized by
\[
\bar{q}_{lm}(i) := \frac{1}{n(i)}\sum_{j=1}^{n(i)} Y_{lm}\left(\vec{r}_{ij}\right)\quad ,
\]
where $Y_{lm}\left(\vec{r}_{ij}\right)$ are the spherical harmonics corresponding to the orientation of the vector $\vec{r}_{ij}$ between hard sphere $i$ and its neighbor $j$ in a given coordinate frame. We consider $l=6$ in order to identify local fcc-, hcp- or rcp-structures. A 13--component vector $\vec{q}_{6}(i)$ is assigned to each hard sphere, the elements $m=-6 \dots 6$  of which are defined as 
\begin{equation}
q_{6m}(i) := \frac{\bar{q}_{6m}(i)}{\left(\sum_{m=-6}^6|\bar{q}_{6m}(i)|\right)^{1/2}} \quad . \label{Defq6q6}
\end{equation}
Two neighbors $i$ and $j$ were regarded as ``bonded'' within a crystalline region, if $\vec{q_6}(i)\cdot \vec{q_6}^*(j) > 0.7$. We define $n_b(i)$ as the number of ``bonded'' neighbors of the $i$th hard sphere. If a hard sphere has more than $9$ bonds we consider it crystalline.

\section*{Dynamic properties in the different pinning scenarios}
\begin{figure}[!ht]
\begin{center}
 \includegraphics[width=\columnwidth]{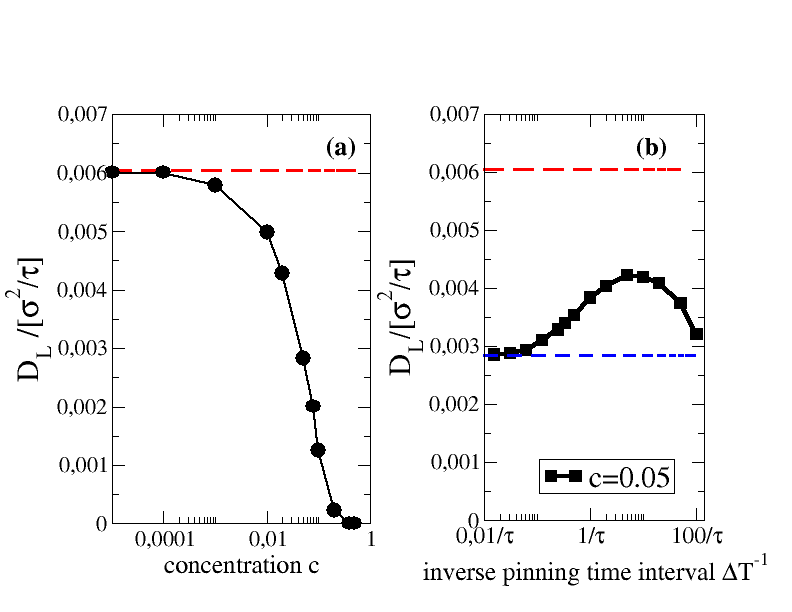}
\end{center}
\caption{\label{fig1} 
 (a) $D_L$ as a function of the static pinning concentration $c$ ($\Delta T\to\infty$). (b) $D_L$ as a function of $\Delta T^{-1}$. The upper dashed lines in the figures indicate the value of $D_L$ at $c=0$ \cite{TaSv2011}. The lower dashed line indicates the value of $D_L$ in the limit of static pinning for $c=0.05$. }
\end{figure}

We start out with the dynamic properties of the supersaturated fluid in the presence of pinned hard spheres. As it has been reported already in \cite{Kim2003,Viramontes-Gambo1995} the overall mobility of the suspension decreases with increasing concentration of pinned hard spheres. We discuss here the diffusion and the relaxation properties. In FIG. \ref{fig1}, the amplitude of the long-time self-diffusion constant $D_L$ is presented. It has been extracted from the mean squared displacement which is defined as  
\[
\lim_{t\to\infty} \langle\Delta r^2(t)\rangle:=\lim_{t\to\infty} \left\langle\frac{1}{N}\sum_{i=1}^N |\vec{ r}_i(t)-\vec{ r}_i(0)|\right\rangle = 6 D_L t~.
\]
The sum includes the mobile as well as the immobile parts of the system. In FIG. \ref{fig1}(a) we present $D_L$ as a function of the static pinning concentration $c$. $D_L$ is decreasing drastically for $c\geq0.01$. For $c \geq 0.2$ the hard spheres cannot explore the entire volume anymore. This leads to an effective diffusion constant $D_L=0$. In FIG. \ref{fig1}(b), for a fixed concentration $c=0.05$, $D_L$ is presented as a function of the inverse of $\Delta T$. Here, dashed lines indicate the diffusion constant of the unpinned system and the diffusion constant for the system at static pinning. For large $\Delta T$, as expected we observe convergence to this value. A maximum is observed around $\Delta T=0.2\tau$ and in the limit of small $\Delta T$, $D_L$ decreases again. Here the pinned hard spheres change very frequently. The resulting dynamics is different from the dynamics of the system without pinning. The diffusive behavior is therefore not expected to be the same. 

\begin{figure}[!ht]
\begin{center}
 \includegraphics[width=\columnwidth]{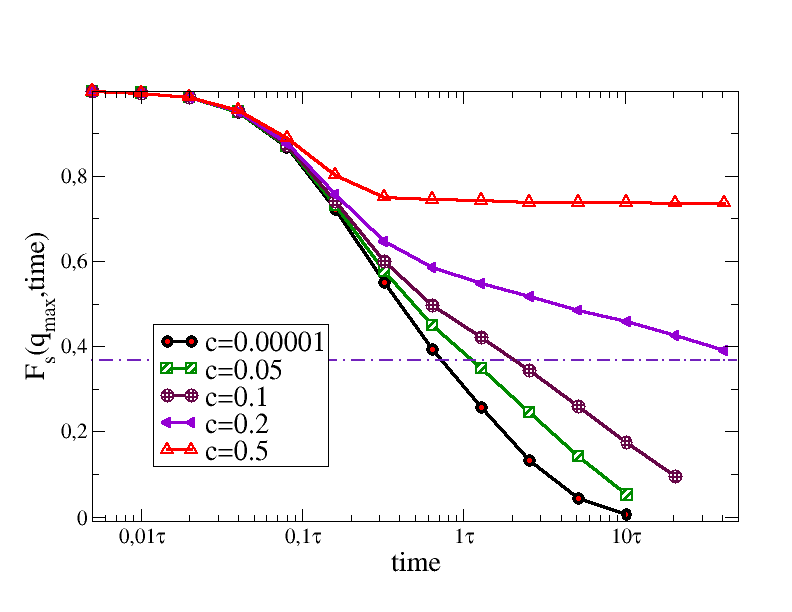}
\end{center}
\caption{\label{fig2}  Dynamic structure factor $F_s(q_{{\rm max}},t)$ for different concentrations $c$ at packing fraction $\eta=0.5393$ for the static pinning scenario. The wave vector amplitude $q_{{\rm max}}$ corresponds to the first peak of the static structure factor $S(q)$. The dashed-dotted line indicates the value $1/e$ ($e$ is Euler's number).}
\end{figure}

In addition to the mean squared displacement, we discuss the properties of the self part of the dynamic structure factor $F_s$, defined as 
\[ 
F_s(q_{{\rm max}},t)=\left\langle\frac{1}{N}\sum_{i=1}^N \exp({\rm i} \vec{q}_{{\rm max}} (\vec{ r}_i(t)-\vec{ r}_i(0)))\right\rangle~.
\]
In FIG. \ref{fig2}, $F_s$ is presented as a function of $c$. 

For large concentrations, $c\geq0.01$, we observe a shoulder as it is characteristic for the slow dynamics in glassy systems \cite{Kim2003, Binder2011}. For the extreme case of $c=0.5$, the dynamic structure factor is not decaying to zero anymore because the mean squared displacement is bounded even for large times.\\
As already mentioned in the introduction, slow relaxation becomes important at high concentrations, but it is still insignificant for concentrations $c\leq 0.05$. 

\section*{Crystallization with static pinning}
Pinning a given concentration of hard spheres introduces a characteristic length scale $l_c$, which interferes with the length associated with the formation of the critical nucleus. Assuming on average an arrangement of the pinned sites in a simple cubic crystal structure, $l_c$ is given by
\[
 l_c= \frac{\sqrt{3}} {\sqrt[3]{c}}\sigma-\sigma.
\]
The diameter of the critical nucleus in the case without pinning at a packing fraction $\eta=0.5393$ is $d_c\approx3.4\sigma$ (which corresponds to approximately 30 hard spheres) \cite{TaSv2011}. 

\begin{figure}[!ht]
\begin{center}
 \includegraphics[width=\columnwidth]{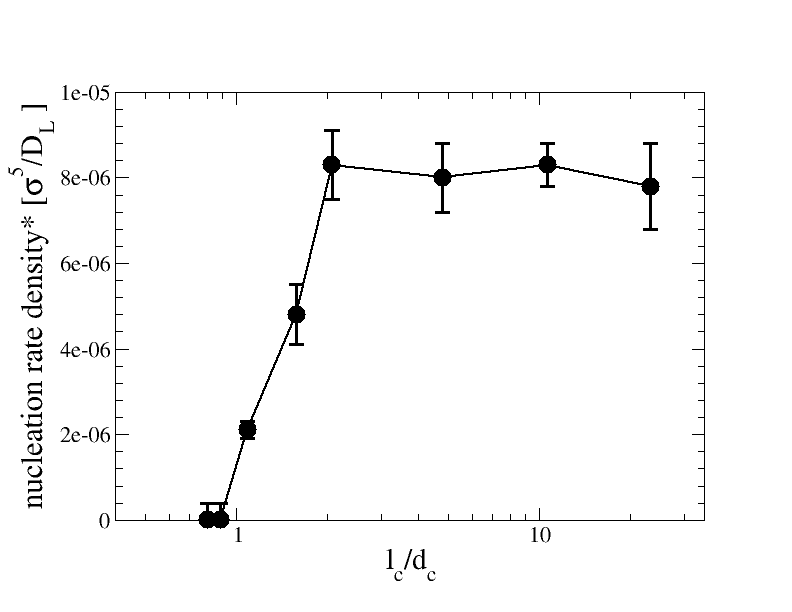}
\end{center}
\caption{ \label{fig5} Nucleation rate densities as a function of the effective average diameter $l_c$ of unpinned regions divided by $d_c$, the diameter of the critical nucleus. Data are compiled from 20 simulation runs each.}
\end{figure}

Crystal nucleation rate densities are presented in FIG. \ref{fig5} as a function of the pinning concentration $c$. We obtain the nucleation rate density as 
\[
 I=\frac{1}{\langle t_c\rangle V}
\]
where $\langle t_c\rangle$ is the mean first passage time to form a stable nucleus and $V$ is the volume of the system.

We observe a sharp decrease in the nucleation rate density around $l_c/d_c\approx1$, which corresponds to a concentration $c\approx 0.05$. When the length scale imposed by the pinned hard spheres becomes smaller than the diameter of the critical nucleus, crystal nucleation is suppressed. \\

\begin{figure}[!ht]
\begin{center}
 \includegraphics[width=\columnwidth]{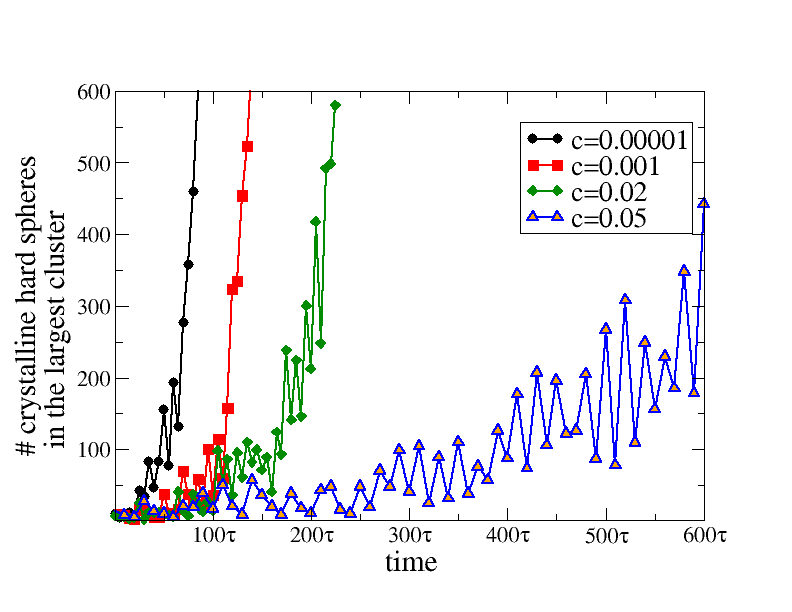}
\end{center}
\caption{ \label{fig4} Time evolution of the number of crystalline hard spheres in the largest cluster as a function of time. Shown are example runs for four different concentrations $c$.}
\end{figure}
\begin{figure}[!ht]
\begin{center}
\includegraphics[width=\columnwidth]{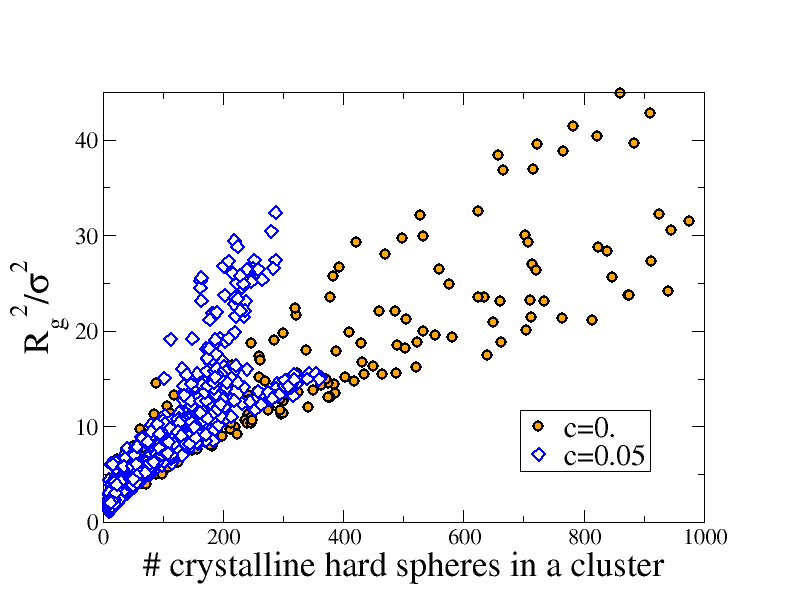}
\end{center}
\caption{ \label{fig7} The radius of gyration $R_g^2$ as a function of the number of hard spheres of high symmetry in a cluster for two different pinning concentrations, $c=0.05$ and $c=0$. Data for 10 simulation runs each.}
\end{figure}

Fig.~\ref{fig4} shows the evolution of the size of the largest cluster for different values of $c$. 
The growth rate is decreasing with increasing $c$ and the size of the cluster as a function of time is fluctuating more strongly.
This indicates internal stresses inside the nucleus, leading to a more irregular structure. This interpretation is supported by the analysis of the radius of gyration $R_g$ as a function of the number of crystalline hard spheres, see FIG. \ref{fig7},
\[
R_g^2=\frac{1}{N_c}\sum_{k,l=1}^{N_c}(\vec{r}_k-\vec{r}_l)^2~.
\]
with $N_c$ the number of hard spheres in a given cluster.\\ 
The radius of gyration of the recorded nuclei for $c=0.05$ is increased compared to the nuclei without pinning. This result clearly shows that the nuclei become more irregular with growing pinning concentration.

\begin{figure}[!ht]
\begin{center}
\includegraphics[width=\columnwidth]{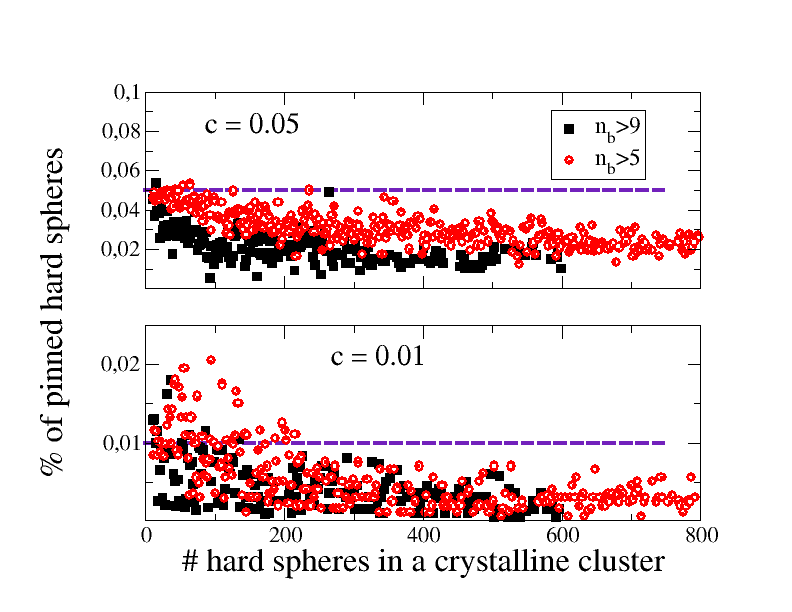}
\end{center}
\caption{ \label{fig8} Mean percentage of pinned hard spheres of high local bond order inside the growing nuclei as a function of the size of the nucleus for $c=0.01$ and $c=0.05$.  The data is averaged over 10 simulation runs each. The dashed lines indicate the system average $c$.}
\end{figure}

We further ask whether pinned hard spheres are part the growing nuclei or whether the nuclei grow such that they avoid them. In FIG. \ref{fig8}, the mean percentage of pinned hard spheres inside the crystalline clusters is recorded as function of the cluster size. 

For the two concentrations $c=0.01$ and $c=0.05$, we note that the percentage of pinned hard spheres inside the crystalline clusters is around half of the system's pinning concentration. (In the case of $c=0.01$, the mean percentage for clusters $N<100$ appears to be greater than the system average because even single pinned hard spheres result in concentrations greater than the system average.) \\
From FIG. \ref{fig7} and FIG. \ref{fig8}, we conclude that the immobile matrix is not incorporated into the growing clusters, which instead become more irregular with increasing $c$. We would also like to point out that our results indicate that single immobile hard spheres do not act as seeds for crystallization. Seeding of crystals requires a larger template, as has also been discussed in \cite{vanBlaaderen, Jungblut}.\\
Our findings motivate the next section of this work, where pinned hard spheres are only held immobile for a given pinning time interval $\Delta T$. 

\section*{Crystallization with periodic pinning}
Static pinning induces defects inside the growing nuclei. We can release the defects on long time scales if we apply periodic pinning, i.e. if a new set of pinned hard spheres is chosen after given time intervals $\Delta T$. We choose a pinning concentration of $c=0.05$ for this analysis.\\

\begin{figure}[!ht]
\begin{center}
\includegraphics[width=\columnwidth]{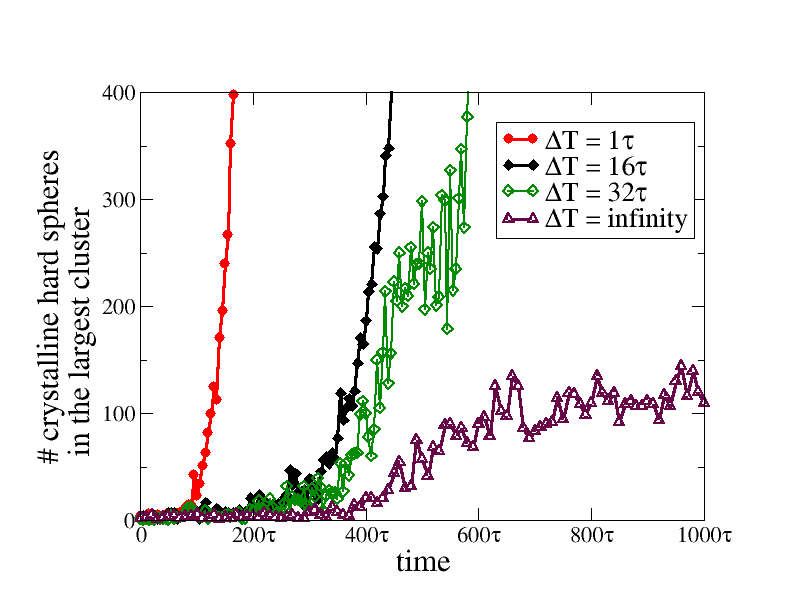}
\end{center}
\caption{ \label{fig9}Time evolution of the number of crystalline hard spheres in the largest cluster as a function of time. Shown are example runs for four different values $\Delta T$ at $c=0.05$.}
\end{figure}
\begin{figure}[!ht]
\begin{center}
\includegraphics[width=\columnwidth]{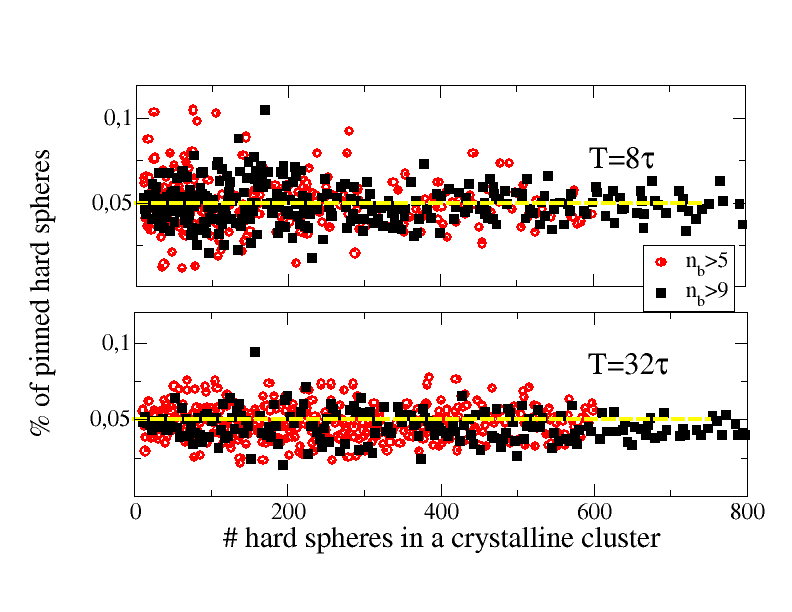}
\end{center}
\caption{ \label{fig10}Mean percentage of pinned hard spheres inside the growing crystal at a pinning concentration $c=0.05$.  The data is shown for two different time intervals $\Delta T=8\tau$ and $\Delta T=32\tau$. 
The data is compiled for 10 simulation runs each. The dotted lines indicate the system average concentration of pinned hard spheres.}
\end{figure}

FIG. \ref{fig9} shows example simulation runs for different values of 
$\Delta T$. We observe that the growth rate is decreasing for large $\Delta T$. Towards the limit of static pinning, see diamond data points for $\Delta T=32\tau$ in FIG. \ref{fig9}, the growing cluster fluctuates strongly in size compared to the smoothly growing clusters for $\Delta T=16\tau$ and $\Delta T=1\tau$. \\
The mean percentage of pinned hard spheres inside the growing crystal is presented in FIG. \ref{fig10}.
We observe that the concentration of pinned hard spheres inside the clusters is equal to the overall pinning concentration $c=0.05$. This leads us to the conclusion that defects are overcome on long time scales. \\

\begin{figure}[!ht]
\begin{center}
\includegraphics[width=\columnwidth]{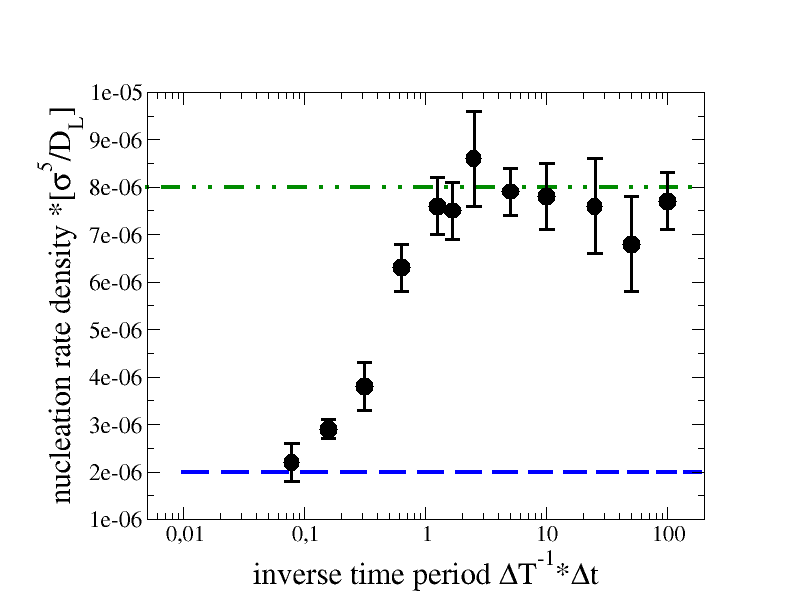}
\end{center}
\caption{ \label{fig6}Nucleation rate densities as a function of the inverse of pinning time interval $\Delta T$ normalized by the typical time a critical cluster needs to develop in the unpinned case, $\Delta t$. The pinning concentration is set to $c=0.05$. The dashed--dotted line indicates the result for the unpinned system, the dashed line indicates the result at static pinning ($\Delta T\to\infty$). Data are averaged over 10 to 20 simulation runs. Error bars indicate the standard deviation.}
\end{figure}

The nucleation rate densities that we obtain for different $\Delta T$ are shown in FIG. \ref{fig6}. 

The time it takes to develop a critical nucleus in the unpinned case is $\Delta t\approx 5\tau$. For $\Delta T\ll\Delta t$, the nucleation rate densities are similar to the unpinned system (indicated by the upper dashed line). However, we have pointed out that the diffusion constant is affected by pinning, i.e the short time dynamics differs between the pinned and the unpinned case. Nevertheless crystallization is not affected as it takes place on time scales that are long compared to $\Delta T$. The system exhibits Brownian motion on this time scale, i.e. the details of the short time dynamics do not matter for the crystallization process \cite{Berthier2007, Patti2012, Sanz2010}.  \\

 For $\Delta T> \Delta t$, we observe a monotonic decrease towards the results of static pinning which is expected in the limit of large $\Delta T$. This limit is indicated by the lower dashed line in FIG. \ref{fig6}. The drop in the nucleation rate density is due to the crossover of time scales. If we pin hard spheres longer than the time it takes for a critical cluster to develop the critical cluster experiences an increasingly static pinning like environment. For small pinning time intervals, all hard spheres in the region where the critical cluster develops move at some point during $\Delta t$ and we recover the nucleation rate density of the unpinned system.

\section*{Conclusion}
We have presented a simulation study of crystallization in suspensions of hard spheres under the constraint of random static and periodic pinning. This approach allows us to directly restrict the number of accessible configurations and the number of possible paths leading to crystallization. \\
We have shown that already a small pinning concentration is sufficient to suppress crystallization completely. We observe a sudden drop in the nucleation rate densities when the length scale introduced by the pinned hard spheres becomes smaller than the diameter of the critical nucleus of the unpinned system (i.e. at a concentration of $c=0.05$ for the supersaturation studied here). 
In addition we characterized the structural properties of the nuclei - more irregular structures are recorded at higher pinning concentrations, the growth rates decrease considerably, and the pinned hard spheres are not incorporated into the crystalline clusters. Since the transition is sharp, one can extract from the measurement of the nucleation rate density at static pinning the typical size of the critical nucleus of the unpinned colloidal system. \\
In a second step we extended the pinning procedure to periodic pinning at a fixed concentration of $c=0.05$. As mentioned earlier $c=0.05$ is close to the sharp decrease in the nucleation rate density. Periodic pinning lifts the internal stresses of the growing nuclei. For small $\Delta T$, the nucleation rate densities approach the one of the unpinned system even though the short time dynamics is different. At $\Delta T > \Delta t$ we observe a monotonic decrease towards the limiting value for static pinning. This decrease is directly related to the time a critical nucleus needs to form in the unpinned system, $\Delta t$. 

The procedure of pinning a low concentration of hard spheres in the overcompressed fluid allows one to obtain information of the static and dynamic properties of the critical nucleus through observing the drop in the nucleation rate densities without exploring the details on the microscopic scale. The results presented here could be experimentally verified for example in colloidal suspensions using laser trapping to pin hard spheres \cite{Cammarota2012} or in two component mixtures with a significant asymmetry in mobility \cite{Krakoviak2010}.

\section*{Acknowledgements}
This project has been financially supported by the DFG (SPP1296) and by the National Research Fund, Luxembourg co-funded under the Marie Curie Actions of the European Commission (FP7-COFUND) and under the project FRPTECD. Computer simulations presented in this paper were carried out using the HPC facility of the University of Luxembourg.


\begin{thebibliography}{999}


\bibitem{Lechner2011} W. Lechner, C. Dellago, P.~G. Bolhuis,
Phys. Rev. Lett., 106, 8, 085701 (2011).

\bibitem{Schilling2010} T. Schilling, H.~J. Sch\"ope, M. Oettel, G. Opletal, I. Snook,
Phys. Rev. Lett., 105, 2, 025701 (2010).

\bibitem{Tanaka2011}H. Tanaka, 
Journal of Physics: Condensed Matter 23, 284115 (2011).


\bibitem{Voivod2009} I. Saika-Voivod, R.~K. Bowles, and P.~H. Poole,
Phys. Rev. Lett., 103:225701 (2009).


\bibitem{Pusey2009} P.~N. Pusey, E. Zaccarelli, C. Valeriani, E. Sanz, W.~C.~K. Poon, and M.~E. Cates, 
Phil. Trans. of the Roy. Soc. A, 367, 1909, 4993-5011 (2009).

\bibitem{Russo2013}J. Russo, A. C. Maggs, D. Bonn and H. Tanaka, Soft Matter 9, 7369 (2013).


\bibitem{Auer2001} S. Auer and D. Frenkel,
Nature 409(6823):1020-3 (2001).

\bibitem{Filion2010} L. Filion, M. Hermes, R. Ni, M. Dijkstra,
The Journal of Chemical Physics, 133, 24, pp. 244115-244115-15 (2010).

\bibitem{TaSv2011} T. Schilling, S. Dorosz, H. J. Sch\"{o}pe, and G. Opletal,
Journal of Physics: Condensed Matter, 23(19):194120 (2011).


\bibitem{Krakoviak2010} V. Krakoviack,
Phys. Rev. E, 82(6):061501 (2010).

\bibitem{Kurzidim2010} J. Kurzidim, D. Coslovich, and G. Kahl,
Phys. Rev. E, 82:041505 (2010).

\bibitem{Kurzidim2011} J. Kurzidim and G. Kahl,
Molecular Physics, 109, 7-10, 1331-1342 (2011).

\bibitem{Kim2003} K. Kim,
EPL (Europhysics Letters), 61(6):790 (2003).

\bibitem{Kim2011} K. Kim and K. Miyazaki, and S. Saito,
Journal of Physics: Condensed Matter, 23(23), 234123 (2011).

\bibitem{Viramontes-Gambo1995} G. Viramontes-Gamboa, J.~L. Arauz-Lara, and M. Medina-Noyola, 
Phys. Rev. Lett., 75, 4, 759--762 (1995).


\bibitem{Scheidler2002} P. Scheidler, W. Kob, and K. Binder,
EPL (Europhysics Letters), 59:701–707 (2002).

\bibitem{Berthier2012} L. Berthier and W. Kob,
Phys. Rev. E, 85:011102 (2012).

\bibitem{Berthier2011} L. Berthier and G. Biroli, 
Reviews of Modern Physics, 83:587–645 (2011).

\bibitem{Krakoviak2005} V. Krakoviack,
Phys. Rev. Lett., 94:065703 (2005).

\bibitem{Krakoviak2011} V. Krakoviack,
Phys. Rev. E 84, 050501 (2011).

\bibitem{Cammarota2012} C. Cammarota and G. Biroli, 
PNAS, 109(23), 8850-8855 (2012).

\bibitem{Lang2010} S. Lang, V. Botan, M. Oettel, D. Hajnal, T. Franosch, and R. Schilling, 
Phys. Rev. Lett. 105, 125701 (2010). 

\bibitem{Alder1959} B. J. Alder and T. E. Wainwright.
The Journal of Chemical Physics, 31(2):459–466 (1959).

\bibitem{Lubachevsky1991} B.~D. Lubachevsky,
J. Comput. Phys., 94:255–283 (1991).

\bibitem{Steinhardt1983} P. J. Steinhardt, D. R. Nelson, and M. Ronchetti,
Phys. Rev. B, 28(2):784–805 (1983).

\bibitem{tenWolde1995} P. Rein ten Wolde, M.~J. Ruiz-Montero, and D. Frenkel,
Phys. Rev. Lett., 75(14):2714–2717 (1995).


\bibitem{Binder2011}		K.~Binder and W.~Kob, 
\emph{Glassy Materials and Disordered Solids: An Introduction to Their Statstical Mechanics, Revised Edition}, World Scientific, Singapore, 2011.




\bibitem{vanBlaaderen}
M. Hermes, E. C. M. Vermolen, M. E. Leunissen, D. L. J. Vossen,
P. D. J. van Oostrum, M. Dijkstra, and A. van Blaaderen, Soft
Matter 7, 4623 (2011).

\bibitem{Jungblut}
S.~Jungblut and C.~Dellago,
Phys. Rev. E {\bf 87}, 012305 (2013).

\bibitem{Berthier2007}		L.~Berthier and W.~Kob, Journal of Physics: Condensed Matter, {\bf 19}, 205130 (2007).
\bibitem{Patti2012}		A.~Patti and A.~Cuetos, Phys. Rev. E, {\bf 86}, 011403 (2012).
\bibitem{Sanz2010}		E.~Sanz and D.~Marenduzzo, The Journal of Chemical Physics, {\bf 132}, 194102 (2010).





\end{thebibliography}

\end{document}